\documentclass[%
 aip,
 amsmath,amssymb,
reprint, %onecolumn %%%%% Comment onecolumn to override one column format %%%%%%%
]{revtex4-1}

\usepackage{graphicx}
\usepackage{dcolumn}
\usepackage{bm}

\usepackage[utf8]{inputenc}
\usepackage[T1]{fontenc}
\usepackage{mathptmx}
\usepackage{etoolbox}

\makeatother
\begin{document}
\newcommand{\ie}{\textit{i.e.,}\ }
\newcommand{\eg}{\textit{e.g.,}\ }
\tolerance=1
\emergencystretch=\maxdimen
\hyphenpenalty=10000
\hbadness=1000

\preprint{AIP/123-QED}

\title[]{Radiation Reaction Kinetics and Collective QED Signatures}
\author{A. Griffith}
\email{arbg@princeton.edu}
\author{K. Qu}
\author{N. J. Fisch}
\affiliation{ 
Astrophysical Sciences, Princeton University}

\date{\today}

\begin{abstract}
  Observing collective effects originating from the interplay between quantum electrodynamics and plasma physics might be achieved in upcoming experiments.
  In particular, the generation of electron-positron pairs and the observation of their collective dynamics could be simultaneously achieved in a collision 
  between an intense laser and a highly relativistic electron beam through a laser frequency shift driven by an increase in the plasma density increase.
  In this collision, the radiation of high energy photons will serve a dual purpose: first, in seeding the cascade of pair generation; and, second, in decelerating the created pairs for detection.
  The deceleration results in a detectable shift in the plasma frequency.  
  This deceleration was previously studied considering only a small sample of individual pair particles.
  However, the highly stochastic nature of the quantum radiation reaction in the strong field regime limits the descriptive power of the average behavior to the dynamics of pair particles.
  Here, we examine the full kinetic evolution of generated pairs in order to more accurately model the relativistically adjusted plasma density.
  As we show, the most effective  pair energy for creating observable signatures occurs at a local minimum, obtained at finite laser field strength due to the tradeoff between pair deceleration and the relativistic particle oscillation at increasing laser intensity.
  For a small number of laser cycles, the quantum radiation reaction may re-arrange the generated pairs into anisotropic distributions in momentum space, 
  although, in the one dimensional simulations considered here, this anisotropy quickly decreases.
\end{abstract}

\maketitle

\section{Introduction}
\label{sec:intro}
Reaching the regime where quantum electrodynamics (QED) influences collective plasma behavior is of wide interest.~\cite{di_piazza_extremely_2012,zhang_relativistic_2020,fedotov_advances_2023}
New interactions between quantum electrodynamics and plasma physics with relevance to extreme astrophysical environments~\cite{sironi_relativistic_2014,turolla_magnetars_2015,uzdensky_extreme_2019} might be explored directly at next generation laboratory experiments.~\cite{chen_perspectives_2023}
High field laser experiments~\cite{chen_relativistic_2009,sarri_generation_2015} will probe the strong field QED regime.
This regime can be characterized by significant nonlinear quantum parameter
\begin{equation}
\chi = \frac{\gamma}{E_c}
\sqrt{\left| \mathbf{E} + c^{-1}\mathbf{v}\times\mathbf{B}\right|^2 - (c^{-1}\mathbf{v}\cdot \mathbf{E})^2}, \label{eq:chi}
\end{equation}
for electron velocity $\mathbf{v}$ with corresponding Lorentz factor $\gamma$ and electromagnetic field $\mathbf{E},\mathbf{B}$ scaled to the Schwinger critical field $E_c$.
Achieving a $\chi \gg 1$ to probe the strong field regime requires a combination of a high strength electromagnetic field and relativistic electrons.
This may be achieved with an all optical approach using the beat wave of two intense lasers,~\cite{zhu_dense_2016,grismayer_laser_2016} or with the collision between an electron beam and a single intense laser.~\cite{abramowicz_letter_2019,yakimenko_prospect_2019,meuren_seminal_2020,meuren_mp3_2021}
The production of high energy photons, through quantum synchrotron radiation, and decay, through the Breit-Wheeler process, should create a cascade of electron-positron pairs.
With a large enough density and volume of electron-positron pairs, experiments can reach the so called QED plasma regime.

Our study aims to characterize the collective behavior of the QED plasma in new experiments.
To experimentally reveal collective effects, the generated pairs need to be sufficiently dense and slow.
These requirements can be met in the collision between a high intensity laser and a high energy electron beam.
In this configuration, the generation of a QED plasma and the observation of collective effects are simultaneously accomplished.
As electrons and positrons increase in density and drop in energy, the driving laser pulse  experiences a frequency shift as the plasma density increases from the added pairs.~\cite{qu_signature_2021,qu_collective_2022}
The frequency shift caused by the creation of new charged particles, analogous to the shift caused by ionization,~\cite{wilks_frequency_1988,joshi_demonstration_1990,savage_frequency_1993,mendonca_theory_2000-1,qu_theory_2018,qu_laser_2019} could serve as an observable signature of collective QED plasma effects.

The role of the pair slowdown -- and of reflection to magnify the contribution of the pair particles -- to producing the frequency signatures was demonstrated in previous work.~\cite{griffith_particle_2022}
This occurs through a combination of two effects.
When pairs have high counter-propagating momenta in the intense laser field such that $\chi \geq 10^{-1}$, the radiation reaction may provide a significant average deceleration.
When the particles are reduced to a low enough energy, the effects of Lorentz force on particle momenta may become significant, and provide a further ponderomotive force in the presence of a laser gradient to temporarily decelerate and even reflect pair particles.
Previous work~\cite{griffith_particle_2022} focussed on the expected mean behavior of pair particles experiencing a radiation reaction slowdown in a strong electromagnetic wave.
However, as the contribution of particles is inversely weighted by their energy;  the most decelerated particles will contribute more significantly.
Thus, the plasma frequency, and thereby the plasma signatures, will be dominated by slow particles.
However, as we address here,  capturing the full dynamics of the low energy tail requires a fuller kinetic picture which captures the stochastic effects of the quantum radiation reaction across the full range of particle momenta.

The quantum radiation reaction's effect on pair behavior is qualitatively different from that of the classical radiation reaction.
The importance of the transition between the two has been demonstrated in great detail.
This occurs as $\chi$ becomes non-negligible.
Theoretical results anticipate discrepancies in emission spectra between classical models, which include the Lorentz-Abraham-Dirac and Landau-Lifshitz, and QED models, which include the Local Constant Field Approximation (LCFA) and the Local Monochromatic Approximation (LMA).~\cite{blackburn_radiation_2020, gonoskov_charged_2022-1,heinzl_locally_2020}
A number of recent executed and proposed experiments have sought to distinguish between these models and validate them within the transition from the classical to the QED regime.~\cite{blackburn_quantum_2014,ridgers_signatures_2017,poder_experimental_2018,cole_experimental_2018,li_angle-resolved_2017,gong_brilliant_2018}
The common experimental configurations used to examine the radiation reaction typically feature an intense laser set against a counter-propagating high-energy electron beam.
This is a similar configuration, but at lower $\chi$, than envisioned for the proposed pair generation experiments.
We aim to address the relevance of changes in the momentum distribution to possible collective signatures.

The different models of the radiation reaction have different implications for plasma kinetics.
In particular, previous work highlighted that, depending on the regime, the radiation reaction may serve to cluster or disperse particles in momentum space.~\cite{lehmann_phase-space_2012,neitz_stochasticity_2013,mackenroth_novel_2013,vranic_quantum_2016,niel_quantum_2018-1}
The stochastic nature of the radiation reaction can allow for qualitatively different particle dynamics.~\cite{geng_quantum_2019,gong_radiation_2019}
In addition to the dynamics of pair particles in an electromagnetic wave, the effect of the radiation reaction on particles in a strong static magnetic field may also shift the momentum distribution.
This has recently been proposed as a possible driver of plasma instabilities.~\cite{bilbao_radiation_2023,zhdankin_synchrotron_2023}
The interplay of the radiation reaction in the QED regime and collective plasma behavior is a developing field, and finding correspondence to detectable signatures in the laboratory would be of great interest.

In this work, we show how signatures of collective effects in the QED plasma regime are affected by the distribution of pair momenta.
We model pairs experiencing a radiation reaction driven deceleration as they transition to the Lorentz force dominated regime.
Particles are initialized at quantum nonlinear parameters higher than previous work to better capture the relevance to experimental configurations which might generate a high density QED plasma.
The full distribution of initially high momentum particles rapidly  slow down and cluster at low momenta in numerical simulations.
Increasing the laser power to decrease the pair momenta results in a higher effective plasma density, until a significant enough intensity is reached, at which point the high oscillatory momentum in the laser dominates and reduces the  signature.
Due to the tradeoff between deceleration and transverse momentum, there is a maximum in signature growth at finite laser intensity.
The distribution of pairs experiences a large initial increase in the spread of momentum in the parallel direction due to the stochastic nature of the quantum radiation reaction.
However, the widening is short lived, as pairs rapidly cluster at low momenta when propagating through the laser pulse.
A large difference in parallel and perpendicular momentum spreads, driven by the radiation reaction, might serve as a source of kinetic instabilities.
In one dimension, for the parameters considered, the radiation reaction cannot serve as a source for a long-lived, highly anisotropic distribution.
However, different pulse configurations or higher dimensional effects could exhibit the increase in the spread of parallel particle momenta.

This paper discusses the distributional dynamics for QED signatures. The paper is organized as follows: 
In Section \ref{sec:eqn} we review the kinetic equation for the radiation reaction in the strong field regime with a  ``collision like'' operator. In Section \ref{sec:num} we detail how this model is numerically applied to the regime of pair slowdown for increasing signatures. In Section \ref{sec:dyn} we evaluate the evolution for frequency signatures.
In Section  \ref{sec:freq} and Subsection  \ref{sec:dist} we examine the relevance of other moments of the distribution.
In Section \ref{sec:disc} we summarize our findings and further discuss the implications of our results.

\section{Distributional dynamic equation}
\label{sec:eqn}
As a basis for determining the dynamics of the pair momentum distribution, we first review a kinetic description of charged particle dynamics including the quantum radiation reaction.
We follow the formalism of Neitz.~\cite{neitz_stochasticity_2013}
The kinetic dynamics are applied to the dynamics of created pairs, where the dynamics are dominated by a strong externally applied laser field, while the fields created by the pair plasma are comparatively weak.
We do not consider the separate problem of pair generation, only radiation and oscillation in the strong externally applied field.

For a homogenous plasma driven by an external laser, the dynamics are described as function of the laser phase $\phi = \omega_0(t-y)$ for a laser propagating in the $y$ direction.
We assume that $t+y$ derivatives vanish, such that there is no spatial dependence to the distribution function $f$ independent of $t-y$.
We note that the distribution of generated pairs, those of primary interest and at low enough energies to be significantly slowed or stopped, 
will be created by the laser, and thus their initial distribution should indeed depend primarily on $\phi$, as creation will be primarily a function of the laser phase.
The external driving laser is taken to have a polarization in the $z$ direction such that the pair $\gamma^2 = 1 + p_y^2 + p_z^2$ and the distribution $f(\phi, p_y, p_z)$ evolves according to the dynamical equation
\begin{equation}
  \frac{p_-}{\gamma}\partial_\phi f + e E(\phi)\left(\frac{p_-}{\gamma}\partial_{p_z}f+\frac{p_z}{\gamma}\partial_{p_y}f\right) = C(f,p_-, \phi).
  \label{eqn:kinetic}
\end{equation}
The momentum coordinate $p_- = \gamma - p_y$ is substituted for convenience as it is conserved by the Lorentz force.
On the right hand side $C$ is a collision like operator coming from the stochastic and non-local emission of high energy photons.
Without further approximation Eqn. \eqref{eqn:kinetic} can be simplified by transforming all derivatives from the $(p_z,\ p_y)$ coordinate system to the $(p_z,\ p_-)$ coordinate system
\begin{equation}
  \partial_\phi f + e E(\phi)\partial_{p_z}f = \frac{\gamma}{p_-}C(f,p_-, \phi).
  \label{eqn:kinetic_simp}
\end{equation}
The transform into the $p_-$ coordinates removes the $\partial_{p_y}$ derivative due to the conservation of $p_-$ by the Lorentz force when the electric field is purely a function of $\phi$.
The left hand side strictly conserves $p_-$ of particles, and the right hand side conserves the $p_z$ of the particles as radiation is taken to strictly occur in the $y$ direction.
Performing a separation of variables on $f(\phi,p_y,p_z) = m(\phi, p_z)n(\phi,p_-)$ a normalized version of $C$ taken to be described by the LCFA~\cite{neitz_stochasticity_2013} may be constructed as
\begin{widetext}
\begin{align}
  \frac{\gamma}{p_-}\frac{2\pi}{\sqrt{3}\alpha |a(\phi)| }\frac{C(f, p_-, \phi)}{m(\phi,p_z)} &=
\int_0^\infty dl\ n(\phi, p\beta)\left[(\beta^{-1}+\beta^{-3}\mathcal{K}_{2/3}(l/\beta) - \beta^{-2}\int_0^\infty dj\mathcal{K}_{1/3}(j+l/\beta)\right]\nonumber \\
 &- n(\phi, p)\int_0^\infty dl \left[(\beta^{-1}+\beta^{-3}\mathcal{K}_{2/3}(l) - \beta^{-2}\int_0^\infty dj\mathcal{K}_{1/3}(j+l)\right].
 \label{eqn:collision}
\end{align}
\end{widetext}
This is written in units where $\hbar = c = m_e = 1$, with a normalized vector potential $\omega_0 a(\phi) = eE(\phi)$ is taken to assume to follow the slowly varying envelope approximation.
Here $\mathcal{K}_\nu$ is the $\nu$'th modified Bessel function of the second kind and for convenience the expression $\beta = 1 + \frac{3}{2}\omega_0 p a(\phi)$ is used.

The kinetic description may be reduced by separation of variables as the distribution function vanishes as the momentum goes to infinity.
The dynamics of $m$ may be found by simply integrating over $p_-$. Given that right hand side redistributes particles in $p_-$ space and conserves particle number it follows that
\begin{equation}
  \partial_\phi m + e E(\phi)\partial_{p_z}m = 0.
  \label{eqn:m_kinetic}
\end{equation}
This equation may be solved through the method of characteristics such that $m\left(\phi, p_z\right) = m\left(\phi_0, p_z + a(\phi)-a(\phi_0)\right)$.
Integrating over $p_z$ gives the separate equation for $n$ which strictly depends on the radiation reaction, to obtain
\begin{equation}
  \partial_\phi n = \frac{\gamma}{p_-}C(n,p_-, \phi).
  \label{eqn:n_kinetic}
\end{equation}
The particular form of $C$ results in a shift towards lower $p_-$ as the distribution evolves in $\phi$.
This operator is non-local in momentum space at significant $\chi$.
As the average $p_-$ decreases, the dynamics become less determined by the stochastic emission of high energy photons, and the radiation reaction should correspond less to the LCFA.
However, the particular choice of model for small $p_-$ (and small $\chi$) over a short duration should not significantly alter the particle dynamics.
Over short times for particles with $p_-$ such that $\chi \ll 1$ the change of the distribution under any model of the radiation reaction should be minimal.

At small $p_-$ when the radiation reaction terminates, the distribution function over $p_-$ will not change significantly, but the distribution over $p_y$ can change significantly.
This can be seen in the map from $p_-$ and $p_y$,
\begin{equation}
  p_y = \frac{1+p_z^2-p_-^2}{2p_-},
  \label{eqn:p_y}
\end{equation}
which picks up a significant nonlinearity when $p_z$ is no longer $\ll p_-$.
As $p_z$ scales with $a_0$, this will occur when particles have been decelerated to energy $\gamma \sim a_0$.
At this energy scale, oscillations of $p_z$ may be strong enough to make $p_y$ go from negative to positive, indicating pair reflection.
For significant $p_-$, reflecting pairs requires a strong laser.
For pairs to be reflected for all time in one dimension $p_- < 1$, which will only be true for a small number of pairs.
However, in higher dimensional simulations, this need not be the case if pairs are scattered to the side, such that they never experience the backend of the laser.
This physics is not captured by Eqn. \eqref{eqn:kinetic}.

In this paper we consider the impact of the radiation reaction not for the purposes of the study of it alone, but for its applicability to frequency shift signatures.
We probe the regime where $\chi \gg 1$ which is expected for the pairs which will be generated in proposed next generation experiments.
Our results differ from those presented in Neitz and Di Piazza,~\cite{neitz_stochasticity_2013} where $\chi$ is increased up to near unity.
Notably, analytical estimates don't capture the pair behavior, as even at moderate $\chi$ large discrepancies emerge between expansions of Eqn. \eqref{eqn:collision} and the full integral expression.
To capture the high $\chi$ regime, the full integro-differential equation must thus be solved, which becomes non-local in momentum space.
To handle the complexity of the expression, this is performed numerically.

\section{Numerical Implementation}
\label{sec:num}

The dynamics of $n$ as described by Eqn. \eqref{eqn:n_kinetic} are evolved numerically using a simple integro-differential solver.
The dynamics are determined for a time-varying scaled laser field which is taken purely to depend on $\phi = \omega(t-x)$, such that there is a perpendicularly polarized $a(\phi)$.
The non-local collision like operator is pre-evaluated at varying $\chi$ using Gauss-Laguerre quadrature.
The term during the dynamical evolution is evaluated over a cubic spline which is fit to the precomputed points which span up to the maximum possible $\chi$ within the simulation.
The time-stepping of $n$ is performed in Julia DifferentialEquation.jl~\cite{rackauckas_differentialequationsjl_2017} at a thousand inhomogeneously distributed points in $p_-$ space.
High momentum packets with an initially Gaussian distribution are initialized within the laser pulse to correspond to generated pairs.
Low $p_-$ points are sampled at higher density to resolve the finer features present at low momenta, which become apparent at the later points in the simulation.
The $p_-$ space is resolved to capture particle momenta ranging from the initial conditions down to the regime where the radiation reaction becomes weak, thresholded numerically by $\chi^2 = (p_-a_0\omega_0)^2 = 10^{-5}$.
Pair-pair interactions, and the effects of the pairs upon the driving laser are not considered beyond the anticipated frequency shift, as both terms are expected and shown to be higher order.

\section{Distributional dynamics for QED signatures}
\label{sec:dyn}
\begin{figure}
  \includegraphics*[width=\columnwidth]{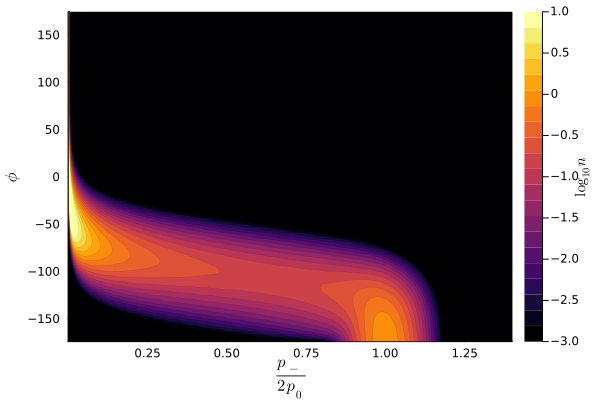}
  \caption{Evolution of $n(p_-)$ in momentum space results in particles moving quickly from high counter-propagating momentum to low counter-propagating momentum.
  At short times the spread in momentum space grows, but as the average particle $\chi$ shrinks, particles quickly cluster in momentum space.}
  \label{fig:density_contour}
\end{figure}

To evaluate total pair contributions we computed the pair momentum distribution evolution in comparison to earlier results on particle slowdown and reflection.
For comparison, pairs are initialized with the same momentum peak in a driving laser with the same envelope and intensity.
For a Gaussian laser pulse of peak intensity $6\times 10^{22}$ W/cm$^2$, particles are initialized in the rising edge with an energy of 10 GeV and an energy spread of 1 GeV.
The dynamics of the distribution in $p_-$ space are shown in Fig. \ref{fig:density_contour}, where particles rapidly slow down and accumulate at low $p_-$.
For initially large $\chi$ due to the emission of high energy photons there is an increase in the spread of energy, however, as the average $\chi$ rapidly decreases the spread correspondingly decreases.

\begin{figure}[t]
  \includegraphics*[width=\columnwidth]{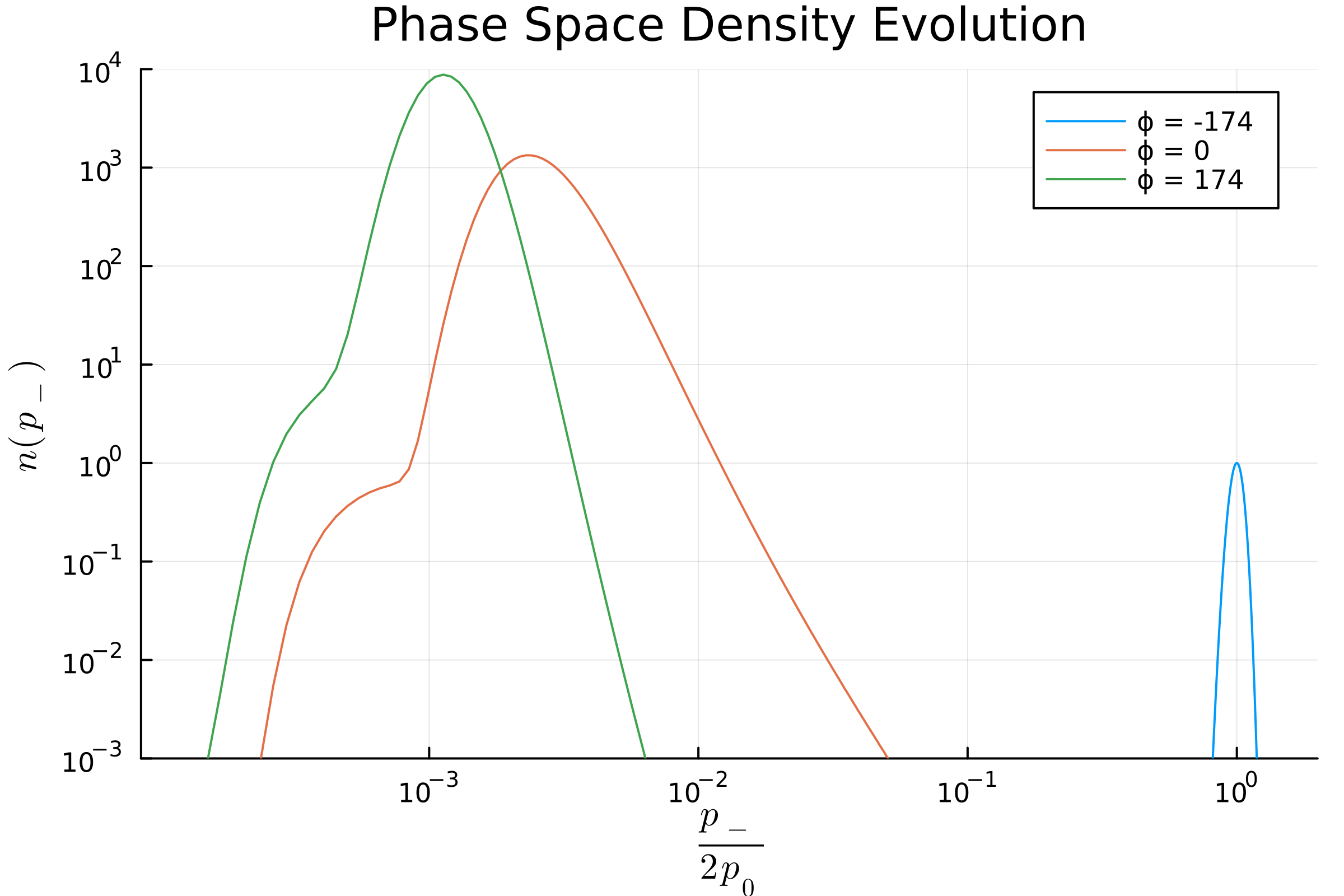}
  \caption{Initial, at focus, and final momentum distributions of particles in the momentum distribution show a significant change in both the spread and mean particle energy at late times.}
  \label{fig:p_m_space}
\end{figure}

While changes in the momentum of the particle distribution function are small compared to the initial energy scale at late times, they are not small compared to their final energy.
This can be seen through a comparison of early, most intense focus, and late times as shown in Fig. \ref{fig:p_m_space}.
Mean $p_-$ still decreases in the back-half of the laser pulse, resulting in an increasing contribution to the frequency signature.
While the spread of the distribution in $p_-$ drops over the duration of the interaction, the spread relative to the mean energy of the particles increases.

\begin{figure}[t]
  \includegraphics*[width=\columnwidth]{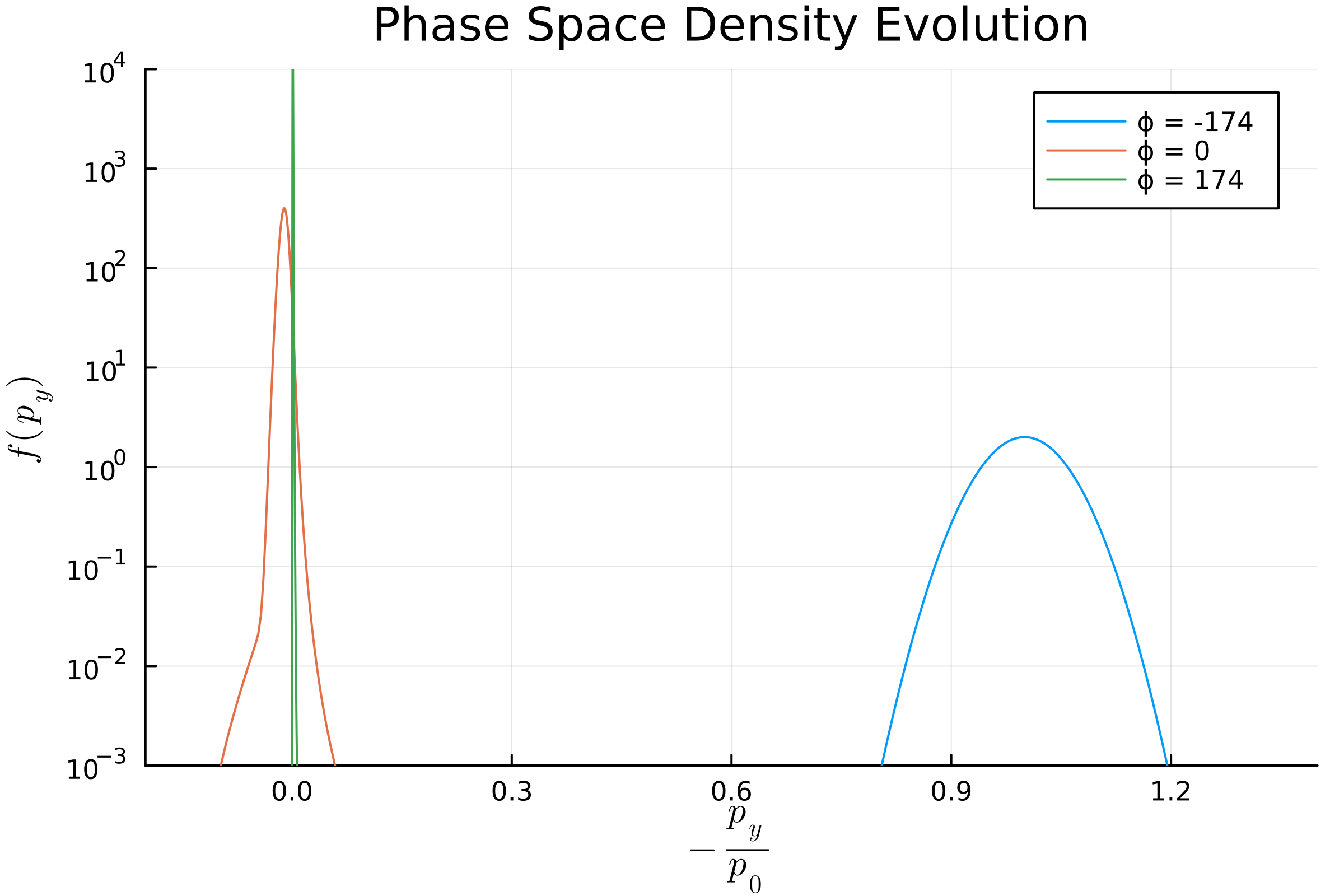}
  \caption{When the pair distribution is scaled from the $p_-$ space to instead be represented in the $p_y$ space, the peaking of the distribution function is sharpened. A significant fraction of particles are reflected at the peak of the laser field, but eventually pairs settle to small, but primarily counter-propagating momenta.}
  \label{fig:p_y_space}
\end{figure}

Changes in the evolution across the $p_-$ axis can be related to the evolution of pairs in $p_y$ space through the conservation of the canonical momentum for a strictly transverse vector potential $\mathbf{A}$.
As pairs progress through the laser field and their energy decreases, eventually oscillations in $p_y$ may become significant when $a(\phi)\sim \gamma$.
This transverse oscillation can be significant enough to induce reflection for a large fraction of the pair particles.
In $p_y$ space over half of the pairs can be temporarily reflected in the peak of the laser pulse as shown in Figure. \ref{fig:p_y_space}.
This requires a significant laser $a_0$, simulations at lower $a_0$ will not decrease $p_-$ to a sufficient enough level for the Lorentz force to overcome the initial particle momenta.
As the laser progresses in one dimension the change in the direction of pairs eventually becomes insignificant as pair particles experience the downramp of the ponderomotive force in the tail end of the laser pulse.

\section{Dynamics of Frequency Shift Signatures}
\label{sec:freq}
The changing momentum distribution of pairs can be directly related to the time varying current in the pair plasma.
The change in the distribution function in $p_-$ alters the frequency of the electromagnetic wave through a time varying susceptibility.
This susceptibility can be found by examining the response of the plasma as determined by the kinetic dynamics of Eqn. \eqref{eqn:kinetic}, which perturbatively changes the electromagnetic wave through an oscillating current
\begin{align}
\Box \mathbf{A} &= \frac{4\pi}{c}\mathbf{J}\\
\mathbf{J} &= e\int d^3p \frac{\mathbf{p}}{\gamma}f.
\end{align}
The oscillations of the pairs are reduced by a factor of $\gamma$ in the relativistic limit.
For a wave polarized in the $z$ direction parallel current, the corresponding $J_z$ is
\begin{equation}
  J_z = e\int_{-\infty}^{\infty} \int_{0}^{\infty} dp_z dp_- \frac{p_z}{\gamma} m(\phi_0, p_z + eA(\phi)-eA(\phi_0))n(\phi, p_-).
  \label{eqn:j_z}
\end{equation}
where the distribution function is represented in $(p_z,\ p_-)$ space and the evolution of $p_z$ is determined by the conservation of canonical momentum.
We note that, in comparison to previous work, density is considered here in the lab frame instead of the pair frame.
When considering many pairs, a single consistent frame must be chosen for pairs of varying momentum. 
The expression given in Eqn. \eqref{eqn:j_z} in the case where the perpendicular momentum spread is small, $m(\phi_0,p_z) = \delta(p_z)$, results in an oscillating transverse current of
\begin{equation}
  J_z =  e^2A(\phi)\int_{0}^{\infty} dp_- \frac{2p_-}{1+a(\phi)^2+p_-^2} n(\phi, p_-),
  \label{eqn:j_z_trans}
\end{equation}
where $\gamma$ has been substituted by the appropriate relation to $p_-$ and $p_z$.
The integral given in Eqn. \eqref{eqn:j_z_trans} serves as the effective plasma density
\begin{equation}
  n_\textrm{eff} = \int \frac{n}{\gamma} dp = \int_{0}^{\infty} dp_- \frac{2p_-}{1+a(\phi)^2+p_-^2} n(\phi, p_-).
  \label{eqn:n_eff}
\end{equation}
The current is suppressed both by a high counter-propagating momentum, and the strong oscillations in the passing wave.
These factors compete depending on the regime.
Initially pairs have $p_- \gg a$, however, $p_-$ may be reduced by the radiation reaction down to the level where they are of similar order.
As the average $p_-$ decreases faster with higher $a$, this results in a balance between which terms dominate within the integral.

\begin{figure}[t]
  \includegraphics*[width=\columnwidth]{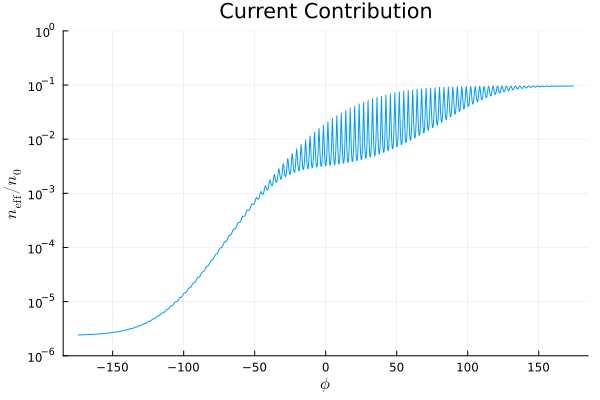}
  \caption{Changes in the $p_-$ momentum distribution increase the effective plasma density and corresponding frequency as pairs slowdown in the laser. The density saturates when oscillations in the transverse momentum become significant enough to impact the average $\gamma$.}
  \label{fig:current_contribution}
\end{figure}

The effective current contribution can be calculated for distribution evolution shown in Figure. \ref{fig:density_contour}.
For a 10 GeV electron originating in the rising edge of an electromagnetic wave at peak intensity of $6\times 10^{22}$ W/cm$^2$ in the rising edge this results in a frequency contribution shift changing with laser phase as shown in Fig. \ref{fig:current_contribution}.
The contributions of pairs rises several orders of magnitude as $p_-$ rapidly decreases.
At low enough $p_-$ the gains of decreased momentum begin to saturate, and pairs oscillate around a slowly decreasing average $\gamma$.
After the peak of the laser pulse, as the laser $a$ decreases, $\gamma$ can continue to decrease as the oscillations become less significant, and a small amount of further decrease in $p_-$ occurs as in Fig. \ref{fig:p_m_space}.

\begin{figure}[b]
  \includegraphics*[width=\columnwidth]{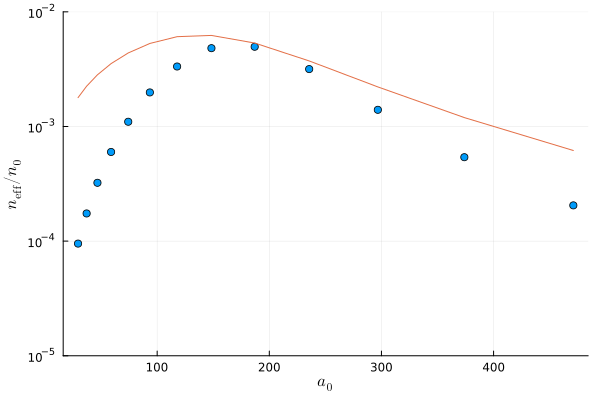}
  \caption{The distribution of pairs is evolved at constant $\chi = p_-(\phi_0) a_0 \omega_0$ for varying $a_0$. Higher $a_0$ induces a larger increase in the effective plasma density for lower $a_0$. However, at high $a_0$, $\gamma$ decreases due to strong transverse oscillations, resulting in a decreasing effective density within the laser pulse.}
  \label{fig:current_contribution_scan}
\end{figure}

The importance of transverse oscillation becomes apparent when considering which configurations of pairs are most effective for generating maximal signature.
If the product of the laser strength and pair momentum are kept constant, such that the initial $\chi$ is constant, pair oscillation may tradeoff against deceleration.
This can be seen in Fig. \ref{fig:current_contribution_scan} where a parameter scan is performed across peak laser $a_0$, keeping $\chi$ constant by decreasing the initial pair momentum.
At low $a_0$ the laser is not significant enough to decelerate particles to a low enough $p_-$ to increase their impact.
At a high $a_0$ transverse oscillations become significant enough to suppress the signal.
The tradeoff between these effects determines the laser power which maximizes the effective density.
A simple model where the radiation reaction is taken to be decelerate particles up to the point where $\chi = 10^{-1}$, such that $p_-(0) = (10\omega_0a_0)^{-1}$ captures the trend of the dynamics.
Assuming that all particles propagate in the laser for enough time to drop to this level predicts an effective density of the form
\begin{equation}
  n_\textrm{eff} \approx n_0 \frac{2 \left(\frac{0.1}{\omega_0a_0}\right)}{1+a_0^2+ \left(\frac{0.1}{\omega_0a_0}\right)^2}.
  \label{eqn:fit}
\end{equation}
This curve is plotted in red in Fig. \ref{fig:current_contribution_scan}, capturing the trend and estimated peak as found in the full kinetic simulations.
This estimate is over-optimistic with regards to both low and high $a_0$, most likely as the $\chi = 0.1$ estimate doesn't capture the radiation reaction dynamics in these limits.
We note that this only optimizes for signature strength given some initial generated pairs.
The efficiency of pair generation, and its dependence on laser strength and the momentum of the electron beam is a separate consideration.
As pair creation will increase with laser field strength, it is expected that the total weighted density will plateau instead of decline.~\cite{qu_signature_2021}
Effective density is of limited use as a single metric for measuring the impact of generated pairs on the laser frequency.
The precise relation between density changes and laser frequency changes instead is determined by a more complicated integral over the time varying density.~\cite{qu_signature_2021}
However, the effective density should serve as a reasonable, simpler proxy for this quantity.

\subsection{Distributional Dynamics}
\label{sec:dist}
To look for possible sources of high anisotropy we characterized the spread of momentum for pairs originating at different points within the laser - electron beam collision.
An increase and then sudden decrease in the spread is notable in Fig. \ref{fig:density_contour}, and we aim to determine for what range of pairs this behavior holds.
It is known that at significant $\chi$ the effective temperature of particles may increase as they stochastically emit photons due to the quantum radiation reaction.~\cite{neitz_stochasticity_2013}
In the evolution of the pair distribution function this can be seen in Fig. \ref{fig:density_contour} where in the rising edge of the laser pulse the distribution greatly widens over the duration of several laser cycles.
However, in the bulk of the laser as the average $p_-$, and correspondingly $\chi$, decreases the growth in the parallel temperature rapidly turns to effective cooling.
A significant increase in the parallel, $p_y$, temperature, with a static perpendicular temperature, $p_z$, could serve to create a large temperature anisotropy.
Temperature anisotropy is a well known source for plasma instabilities.~\cite{weibel_spontaneously_1959,fried_mechanism_1959-1,davidson_nonlinear_1972}
We characterized wether the quantum radiation reaction could serve as a unique source for instabilities.

\begin{figure}[t]
  \includegraphics*[width=\columnwidth]{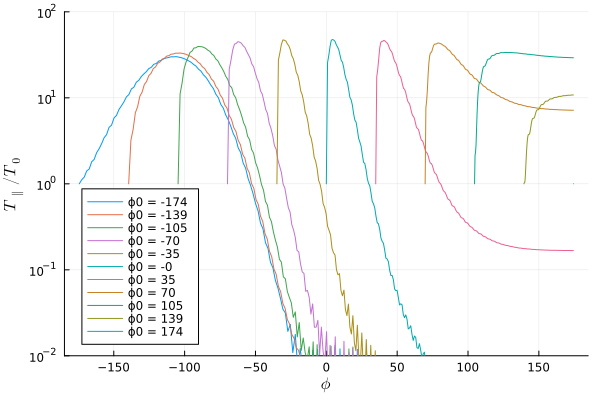}
  \caption{The pair $\langle (p_y - \bar{p_y})^2 \rangle$ initially increases regardless of where pairs are injected in the laser. However, as pairs transition to low momentum and low $\chi$, this quantity rapidly shrinks. Only pairs which originate in the weak tail of the laser pulse maintain a large spread in $p_y$.}
  \label{fig:sigma_p_y_scan}
\end{figure}

Significant increases in pair parallel, relative to the laser propagation direction, temperature cannot be sustained within the peak of the driving laser.
This can be seen in Fig. \ref{fig:sigma_p_y_scan}, where the distribution of pairs is initialized at different phase (times) within the pulse.
The second moment of the distribution is calculated in comparison to the initial conditions
\begin{align}
  \frac{T_\parallel}{T_0} &= \frac{n_0^{-1}\int p_y^2 n(\phi) dp_- - \left(n_0^{-1}\int p_y n(\phi) dp_-\right)^2}{n_0^{-1}\int p_y^2 n(\phi_0) dp_- - \left(n_0^{-1}\int p_y n(\phi_0) dp_-\right)^2},\\
  n_0 &= \int n(\phi) dp_-
\end{align}
Pairs which experience the high intensity parts of the pulse rapidly slowdown and cool, as the field strength is high enough that the radiation dominates.
At low enough $p_-$ oscillations in $p_z$ begin to impact the map from $p_-$ to $p_y$,
even though the distribution is still strictly converging in $p_-$.
Only pairs that are initialized in the tail end of the laser pulse experience the growth in the temperature associated with high $\chi$, without the following fall.
Pairs will primarily be generated in regions of high field, so this suggests that in a full simulation coupling both pair generation and radiation, it would be unlikely for a significant increase in temperature to be observed across all pairs.

High parallel temperatures, and thus high temperature anisotropies, could be maintained in the case where particles exit the high intensity laser field before settling to low energy.
This is present in pairs with originate in the back half of the laser pulse.
Furthermore, if pairs which originate in the rising edge or peak of the laser are quickly side scattered, decreasing the high field they experience, then the high parallel spread in momentum might still remain ``imprinted'' on the distribution.
This effect would not be captured by the current simulation, which is strictly performed in one dimension.
Side scattering into the perpendicular direction has been noted to be significant.~\cite{griffith_particle_2022}
Through these two mechanisms, it is possible that the unique properties of the distribution function at short times could be preserved for observation, or as a source of collective plasma behavior.

\section{Summary and Discussion}
\label{sec:disc}
In conclusion, we provide a full description of pair momentum evolution,  identifying the tradeoff between deceleration and transverse oscillation.
Radiation reaction driven pair deceleration is necessary to produce detectable frequency shifts from electron-positron pair creation.
For the parameters considered in previous work,~\cite{qu_collective_2022,griffith_particle_2022} an intense laser can raise the relativistic mass weighted plasma density orders of magnitude.
As the average counter-propagating momentum, $p_- = \gamma - p_y$, decreases in the driving laser from the radiation reaction, the Lorentz force begins to significantly impact the effective density.
The effect of the Lorentz force is not primarily the redirection of pair particles, but instead an increase in the pair transverse momentum, $p_z$.
When $p_z$ is on the same order of $p_-$, the effective density is reduced from transverse oscillations, inducing a form of relativistic transparency.
This creates a tradeoff between high laser intensity to achieve a rapid deceleration of generated pairs, and a low laser intensity to reduce transverse oscillation.

A naive model for the particle energy, which sets a threshold of $\chi =0.1$ for the radiation reaction, captures the qualitative features of this tradeoff, as shown in Fig. \ref{fig:current_contribution_scan}.
A laser peak of strength $a_0 \sim 150$ produces a maximal contribution of pairs to the frequency signature, assuming a constant initial $\chi$ for the generated pairs.
In isolation from the coupled pair creation problem, this suggests that there is limited benefit from going to higher intensity if the aim is to produce a more distinct signature.
Well known relativistic behavior in plasma physics restricts the observability of the new QED plasma behavior which is of interest.
However, it is worth noting that the full coupled problem of pair creation and deceleration may benefit from a more significant $a_0$.
The tradeoff observed for pair deceleration shows that generation cannot be considered in isolation, and maximizing signatures will require solving the fully coupled problem.

Changes in the width of the momentum distribution are large over short timescales, but are quickly quenched in the high $\chi$ limit.
Thus, the changing momentum distribution alone cannot provide a new source for collective QED plasma behavior.
It is known that the radiation reaction can seed plasma instabilities by creating unstable plasma distributions.~\cite{bilbao_radiation_2023, zhdankin_synchrotron_2023}
In the high $\chi$ regime, when pairs  strictly propagate in one dimension, the distribution function rapidly expanded and then contracted in parallel momentum.
For pairs which originate with a narrow spread of perpendicular momentum, it might be the case that a large spread of parallel momentum could serve as a source for an anisotropic, and thus unstable, distribution.

However, a highly anisotropic momentum distribution cannot be maintained for pairs which experience a multi-cycle laser pulse with high initial $\chi$.
The initially highly stochastic emission of photons at high $\chi$ quickly collapses to low momentum as particles lose energy in the strong laser field.
For strongly anisotropic distributions to exist after interacting with the driving laser pulse, either a distinctly different energy and intensity regime must be considered, or pairs must be pulled out of the laser pulse before they evolve to lower and more tightly clustered momenta.
The latter case of brief exposure to high $\chi$ cannot be achieved in one dimension as all particles pass through the entire wave.
However, it can occur in a two-dimensional laser pulse where side-scattering becomes significant.
In a two or three dimensional simulation, kinetic effects might better correspond to previous work~\cite{qu_collective_2022,qu_pair_2023}, and could offer an explanation for the filamentation therein.

Simulations evolved in laser phase and one momentum dimension serve as a limited starting point in looking for kinetic features in QED experiments.
The addition of spatial variation as well as non-separable momentum coordinates could open up new, richer behavior.
This could be directly addressed through immediate extension of current work to add additional transverse and longitudinal components the distribution function in both space and momentum.
A fuller description might lead to the maintenance of kinetic features for longer durations or allow qualitatively new behavior to develop.
To better understand the stability of QED plasmas, numerically generated distributions can be analyzed for the presence of unstable modes, and the direct study of kinetic instabilities could be examined over longer time scales.

\begin{acknowledgments}
  This research was supported by NSF PHY-2206691.
\end{acknowledgments}

\bibliography{qed_disp}

\end{document}